\documentclass[twocolumn,aps,prl,10pt]{revtex4-1} 

\usepackage{graphicx}
\usepackage{bm}
\usepackage{dcolumn}
\usepackage{amssymb}
\usepackage{amsmath}
\usepackage{amsbsy}
\usepackage{amsfonts}
\usepackage{epsfig}
\usepackage{graphicx}
\usepackage{multirow}
\usepackage{stackrel}
\usepackage{paralist}
\usepackage[rgb]{xcolor}
\usepackage[textsize=tiny,textwidth = 1.6 cm]{todonotes}

\usepackage[percent]{overpic}
\usepackage{hyperref}
\hypersetup{colorlinks=true,allcolors=blue}
\usepackage{tikz}
\usetikzlibrary{intersections} 
\usetikzlibrary{positioning,calc}
\usepackage[caption=false]{subfig}
\usepackage[normalem]{ulem}

\tikzset{
    right angle quadrant/.code={
        \pgfmathsetmacro\quadranta{{1,1,-1,-1}[#1-1]}     
        \pgfmathsetmacro\quadrantb{{1,-1,-1,1}[#1-1]}},
    right angle quadrant=1, 
    right angle length/.code={\def\rightanglelength{#1}},   
    right angle length= 1 ex, 
    right angle symbol/.style n args={3}{
        insert path={
            let \p0 = ($(#1)!(#3)!(#2)$) in     
                let \p1 =  ($(\p0)!\quadranta*\rightanglelength!(#3)$),
                \p2 = ($(\p0)!\quadranta*\rightanglelength!(#2)$) in 
                let \p3 = ($(\p1)+(\p2)-(\p0)$) in  
            (\p1) -- (\p3) -- (\p2)
        }
    }
}

\newcommand{\cancel}[1]{}
\newcommand{\new}[1]{#1}

\newcommand{\I}{\mathrm{i}}

\newcommand{\refEq}[1]{Eq.~(\ref{#1})}
\newcommand{\refFig}[1]{Fig.~\ref{#1}}

\newcommand{\refTab}[1]{Tab.~\ref{#1}}
\newcommand{\citeRef}[1]{Ref.~[\onlinecite{#1}]}

\definecolor{NewColor}{rgb}{1,0,0}

\graphicspath{{../Graphics/}{../Graphics/Publication/}}

\newcommand{\CO}[1]{\textcolor{red}{}}

\newcommand{\expstyle}{\scriptstyle}

\newlength{\noLengthl}

\renewcommand{\vec}[1]{{\boldsymbol{#1}}}

\begin{document}

\title{\texorpdfstring{Winding up quantum spin helices: \\ How avoided level crossings exile classical topological protection}{Winding up quantum spin helices: How avoided level crossings exile classical topological protection}
}
\author{Thore Posske}
\author{Michael Thorwart}
\affiliation{I. Institut f{\"u}r Theoretische Physik, Universit{\"a}t Hamburg, Jungiusstra{\ss}e 9, 20355 Hamburg, Germany}
\begin{abstract}
A magnetic helix can be wound into a classical Heisenberg chain by fixing one end while rotating the other one. 
We show that in quantum Heisenberg chains \new{of finite length}, 
the magnetization slips back to the trivial state beyond a finite turning angle.
Avoided level crossings  
thus undermine classical topological protection.
Yet, for special values of the axial Heisenberg anisotropy, stable spin helices form again, which are non-locally entangled. Away from these sweet spots, spin helices can be stabilized dynamically or by dissipation. For half-integer spin chains of odd length, a spin slippage state and its Kramers partner define a qubit  with a non-trivial Berry connection.
\end{abstract}

\maketitle

The ongoing downsizing of bits in computational devices is about to hit the limits given by the coarse-grained character of matter. Among the promising candidates to store information on the atomic scale are non-collinear magnetic structures like domain walls, spin helices, and magnetic skyrmions \cite{Parkin..Thomas2008MagneticDomainWallRacetrackMemory,Tomasello..Finocchio2014AStrategyForTheDesignOfSkyrmionRacetrackMemories,Hagemeister..Vedmedenko..Wiesendanger2016SkyrmionsAtTheEdge:ConfinementEffectsInFeIr111}.
These kinds of magnetic structures cannot be continuously deformed to a trivial state, e.g., a \new{ferromagnetically ordered} one, without letting the magnetization vanish at some point. The information is topologically protected.
However, systems with a few spins have two limitations. Classical topological protection usually decreases with size because the system is more discrete than continuous. Moreover, small systems are inherently governed by quantum effects. Topologically protected states may be left by tunneling. 

Conceptually simple magnetic structures with topological protection are classical magnetic helices.
It is well-established in the context of a spin energy storage, that
the magnetization of classical
spin chains of finite length can be wound up to a helix when the first spin is rotated slowly while the last one is fixed \cite{VedmedenkoAltwein2014TopologicallyProtectedMagneticHelixForAllSpinBasedApplications}.
\cancel{This could be achieved by locally controllable magnetic moments of magnetic islands  \cite{Khajetoorians..WiebeWiesendanger2013CurrentDrivenSpinDynamicsOfArtificiallyConstructedQuantumMagnets,KhajetooriansWiebe..Wiesendanger2011RealizingAllSpinBasedLogicaOperationsAtomByAtom}.}
\cancel{Depending on the length of the chain,}
A finite number of rotations is possible before the spins slip back %
and the system partially releases its attained energy. 
Besides acting as a spin energy storage, by tuning the winding number of a spin helix, full control of the overlap between the Majorana bound states of helically magnetized one dimensional topological superconductors \cite{Kjaergaard..Flensberg2012MajoranaFermionsInSCNanowiresWithoutSOC} becomes possible. This implements additional  \cite{AliceaOregRefaelVonOppenFisher2011NonAbelianStatisticsAndTopologicalQuantumInformationProcessingIn1DWireNetworks} dynamical quantum gates.
Static helices that rely on RKKY or Dzyaloshinskii-Moriya interactions do not offer this possibility \cite{Klinovaja..Loss2013TopologicalSuperconductivityAndMajoranaFermionsInRKKYSystems,BrauneckerSimon2013InterplayBetweenClassicalMagneticMomentsAndSCInQuantum1DConductors,VazifehFranz2013SelfOrganizedTopologicalStateWithMajoranaFermions}.

The extension of the concept of spin helix states to the quantum regime may evoke surprising new properties of which a few have been \new{recently revealed theoretically}
\cite{Popkov2013PhysRevE,Popkov2016PhysRevA,Popkov2017JPhysA,PopkovSchuetz2017SolutionOfTheLindbladEquationForSpinHelixStates,Popkov2017PhysRevA}.
Stable quantum spin-{\small$1/2$} helices at infinitely strong coupling of the first and last spin of the chain to dissipative baths, for instance, 
have been shown to exist only for specific values of the axial Heisenberg anisotropy matching the cosine of the relative turning angle between two neighboring spins \cite{Popkov2016PhysRevA,Popkov2017JPhysA}. 
At these fine-tuned sweet spots, the helix state is a pure product state of local spin states \cite{Popkov2017JPhysA,Popkov2017PhysRevA, PopkovSchuetz2017SolutionOfTheLindbladEquationForSpinHelixStates}.
Spin helices and spin slips additionally appear in superfluid spin transport  
and are related to superconducting charge transport in thin wires
\cite{Sonin1978AnalogsOfSuperfluidCurrentsForSpinsAndElectronHolePairs,Sonin2010SpinCurrentsAndSpinSuperfluidity,KoenigBonsagerMacDonal2001DissipationlessSpinTransportInThinFilmFM,ChenSigrist2014SpinSuperfluidityInCoplanarMultiferroics,KimTakeiTserkovnyak2016ThermallyActivatedPhaseSlipsInSuperfluidSpinTransportInMagneticWires,KimTserkovnyak2016TopologicalEffectsOnQuantumPhaseSlipsInSuperfluidSpinTransport,Little1967DecayOfPersistentCurrentsInSmallSuperconductors,LangerAmbegoakar2967IntrinsicResistiveTransitionInNarrowSCChannels,McCumberHalperin1970TimeScaleOfIntrinsicResistiveFluctuationsInThinSCWires}.
\new{
Experimentally, few-atom spin chains are at the forefront of research, realizing locally controllable magnetic moments of magnetic islands \cite{Khajetoorians..WiebeWiesendanger2013CurrentDrivenSpinDynamicsOfArtificiallyConstructedQuantumMagnets}, boundary-controlled spin manipulation \cite{ KhajetooriansWiebe..Wiesendanger2011RealizingAllSpinBasedLogicaOperationsAtomByAtom}, and noncollinear magnetism \cite{Steinhoff..Wiesendanger2012IndividualAtomicScaleMagneticsInteractingWithSpinPolarizedFieldEmittedElectrons}.
In these setups, the finite size of the spin chains is crucial regarding the observed physics and prospects for computational applications.
}

In this Letter, we show that the dynamic winding-up of Heisenberg quantum spin chains \new{of finite length} reveals non-trivial quantum mechanical features. First, quantum spin slippage occurs in the winding process which is absent in classical chains and which prevents stable quantum spin helices from occurring. Generic quantum-mechanical avoided energy level crossings let the quantum spin chain prematurely slip to the trivial collinear state. 
Second, we find for general spin quantum numbers a cascade of sweet spots of the axial anisotropy for which most relevant avoided level crossings numerically vanish and which include the special cases for the quantum spin-{\small$1/2$} chains with infinitely strong boundary dissipation as a subclass \cancel{Ref.} \cite{Popkov2017JPhysA}. Third, we find that the quantum spin helix state is non-locally entangled, which generalizes \citeRef{Popkov2017JPhysA}. 
In addition, we show that \new{finite-size} quantum spin helices can, in general and away from the sweet spots, be realized by dynamic winding protocols that exploits Landau-Zener transitions, or by coupling all spins to a dissipative magnetic environment.
Furthermore, we point out that the quantum slippage states themselves are interesting objects: For chains of half-integer spins with an odd length, the energetically lowest slippage state is a Kramers partner of the ground state at a twisting angle of $\pi/2$ and both states are separated energetically from the rest of the spectrum. These states define a qubit with a nontrivial gate operation realized by adiabatic time evolution. 

\begin{figure*}
\subfloat[\label{figClassicalSpinSpiralling}]{%
\raisebox{-0.5 \height}{\includegraphics[height = 0.175 \linewidth]{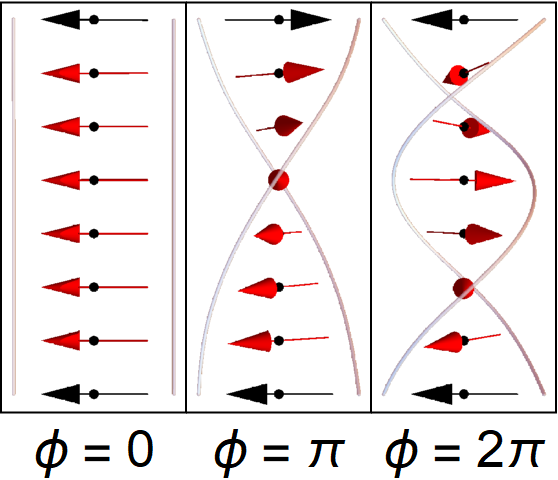}
}
\raisebox{-0.495\height}{\includegraphics[height = 0.175 \linewidth]{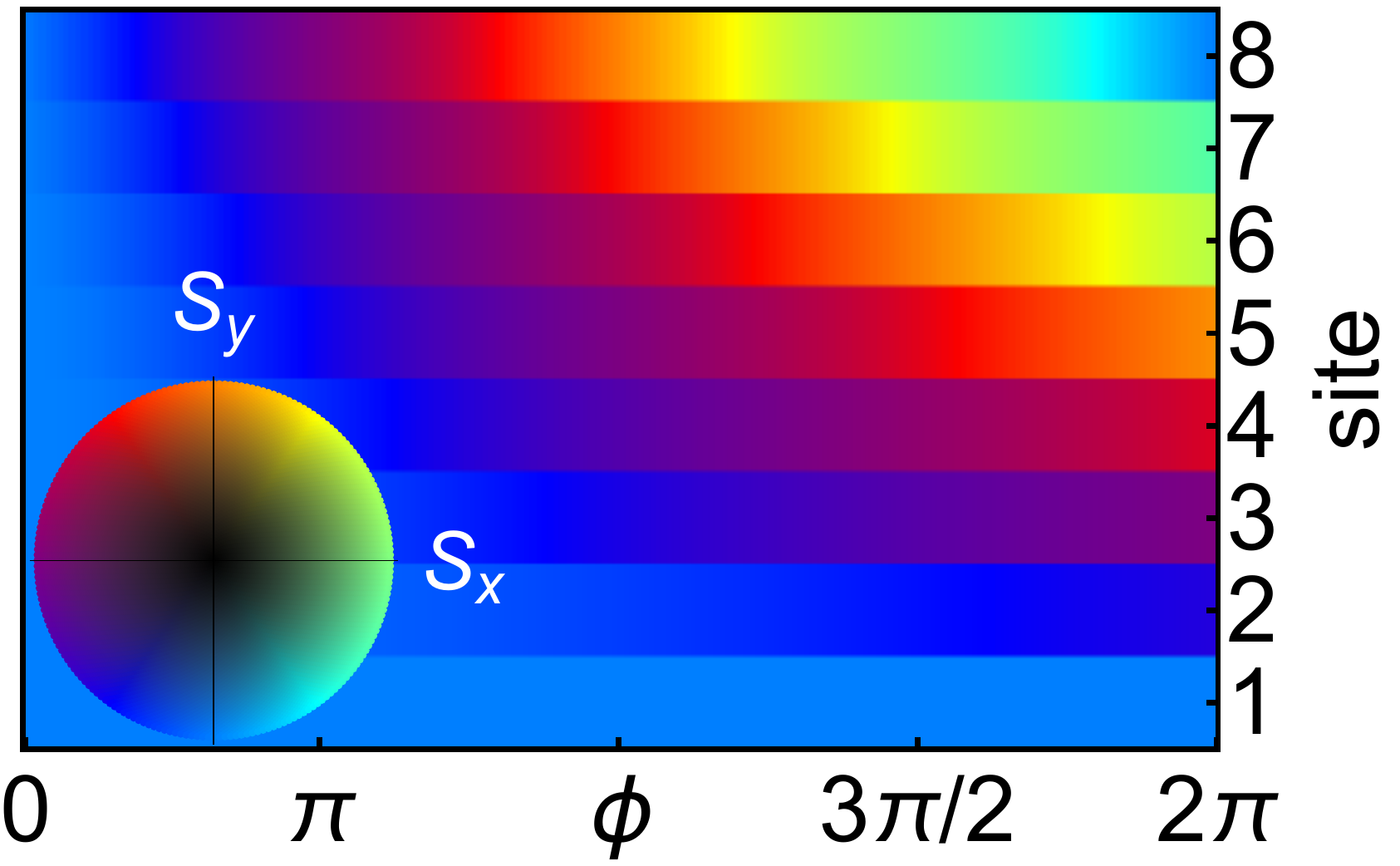}%
}}
\hfill
\subfloat[\label{figQuantumSpinSpiralling}]{%
\raisebox{-0.5 \height}{\includegraphics[height = 0.175 \linewidth]{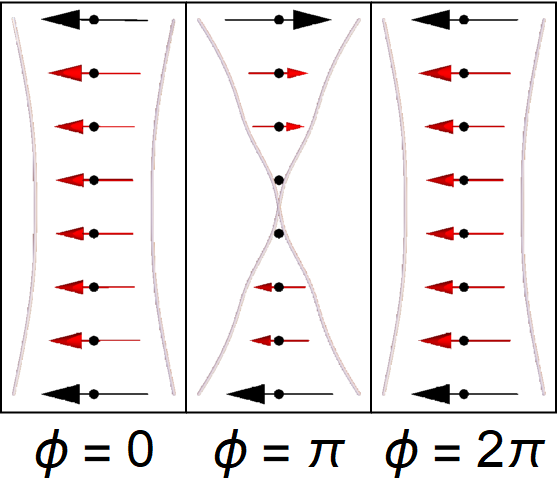}}%
\raisebox{-0.495\height}{\includegraphics[height = 0.175 \linewidth]{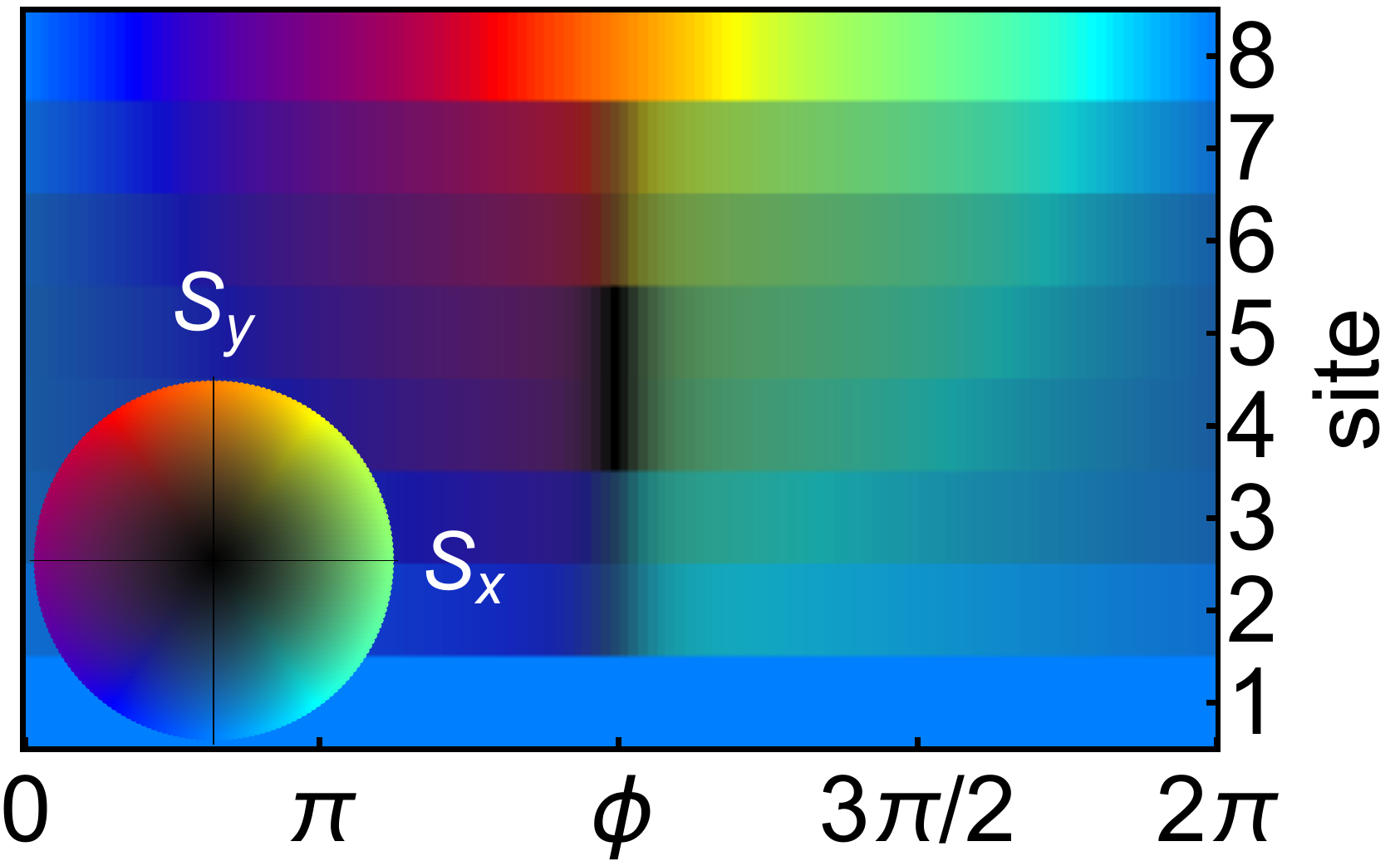}}
}
\caption{\label{figUpWindingOfSpinSpirals}%
Spin orientation during winding up spin helices.
The first spin is fixed while the last one is adiabatically slowly rotated by the angle $\phi$. \textbf{a)} Classical spins, simulated with the Landau-Lifshitz-Gilbert equation corresponding to \refEq{eqnHamiltonian} and large Gilbert damping. A spin helix develops. \textbf{b)} Quantum spins, shown are the expectation values $\langle \vec{S}_i\rangle$ at $\Delta=0$. At $\phi=\pi$, the middle spins slip where the local spin expectation reaches zero. No helix develops.
}
\end{figure*}
%
\paragraph{Model. }
We consider the $XXZ$ Heisenberg model of a chain of $n$ quantum spins $\vec{S}_i$ 
 \cite{KirillovReshetikhin1986ExactSolutionOfTheHeisenbergXXZModelOfSpinS,Nepomechie2001BetheAnsatzSolutionOfTheOpenXXSpinChainWithNondiagonalBoundaryTerms,LuMullerKarbach2009QuasiparticlesInTheXXZModel}. 
The terminal spins are completely fixed by external control fields, for instance, stemming from magnetic islands as experimentally realized in \citeRef{KhajetooriansWiebe..Wiesendanger2011RealizingAllSpinBasedLogicaOperationsAtomByAtom}.
Effectively, the terminal spins can be treated classically.  
The Hamiltonian is 
\begin{align}
\label{eqnHamiltonian}
\mathcal{H}(t) =&\sum_{i=2}^{n-2} J 
	\left(
			 {S_i^x  S_{i+1}^x +  S_i^y S_{i+1}^y}
	\right) 
	+ \Delta S_i^z S_{i+1}^z \nonumber\\& 
	+ J S \left( \vec{\hat{B}}_1 \vec{S}_2 + \vec{\hat{B}}_n(t) \vec{S}_{n-1}\right).
\end{align}
Here, $J$ is the Heisenberg exchange coupling and $\Delta$ the axial Heisenberg anisotropy.
The external field $\vec{\hat{B}}_1 = \left(1,0,0\right)^T$ fixes the first spin of the chain, the field $\vec{\hat{B}}_n(t) = \left(\cos \phi(t) , \sin \phi(t) , 0 \right)^T$ rotates the last spin, e.g., $\phi(t) = \omega t$. 
\new{
The finite size of the spin chain here plays the important role of letting the external fields polarize the center of the chain significantly. In the thermodynamic limit, i.e., $n\to \infty$, the ground state can be unordered and gapless depending on $J$, $\Delta$, and the spin quantum number \cite{GiamarchiBook2003}.
}
For simplicity, we assume a ferromagnetic coupling, i.e., $J, \Delta< 0$, and a preferred orientation of the spins in the $x$-$y$ plane, i.e., $|\Delta|<|J|$.

\paragraph{Quantum spin slippage.}
Initializing the spin chain in its \new{ordered} ground state and letting time run, the last spin is rotated. %
For sufficiently slow dynamics, the spins gradually follow the orientation of their nearest neighbors. In a classical chain, a spin helix develops, as shown in \refFig{figClassicalSpinSpiralling}. 
For quantum spins \new{in finite chains}, however, the situation is different. In \refFig{figQuantumSpinSpiralling}, we depict the expectation values of the spins $\langle \vec{S} \rangle$. Instead of forming a helix, the expectation values of some spins vanish at a twisting angle around $\pi/2$. This behavior appears for almost all values of $\Delta$ in the planar regime. We denote this phenomenon as quantum spin slippage.

\begin{figure*}
\subfloat[\label{figTimeEvolvedSpectrumNotSweet}]{%
\includegraphics[width = 0.25 \linewidth]{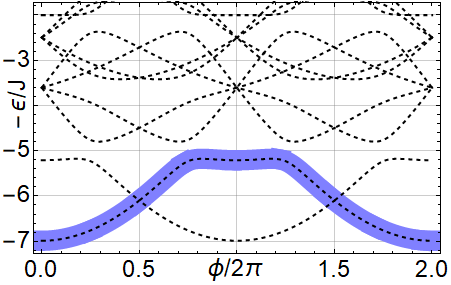}}%
\subfloat[\label{figTimeEvolvedSpectrumNotSweetDiabatic}]{%
\includegraphics[width = 0.25 \linewidth]{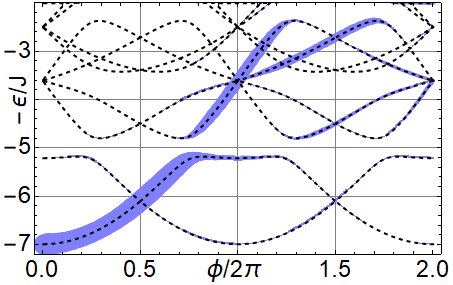}}%
\subfloat[\label{figTimeEvolvedSpectrumSweet}]{%
\includegraphics[width = 0.25 \linewidth]{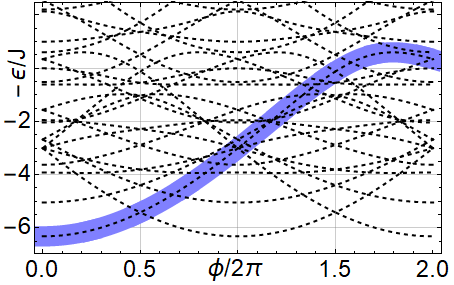}}%
\subfloat[\label{figTimeEvolvedSpectrumSweetDiabatic}]{%
\includegraphics[width = 0.25 \linewidth]{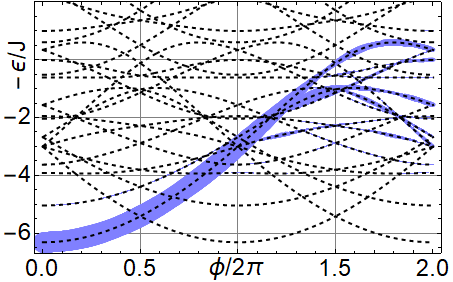}}%
\caption{\label{figTimeEvolvedSpectrum}%
Adiabatic and diabatic time evolution of the ground state. The instantaneous spectrum, %
\refEq{eqnHamiltonian}, is marked by dotted black lines, the weight of the time-dependent state projected onto the instantaneous eigenstates is reflected by the width of the blue stripes.  
\textbf{a)} Adiabatic evolution for $\Delta = 0$. The two energetically lowest states are separated from higher states by an avoided level crossing. No helical state develops.
\textbf{b)} Diabatic evolution for $\Delta = 0$. The avoided level crossing is not perfectly overcome by rotating the spins with a finite angular velocity, $\phi(t) = -0.3 \new{J t /\hbar}$.
\textbf{c)} Adiabatic evolution for the sweet spot $\Delta = J/2$. The avoided level crossings close. 
A helical state develops. 
\textbf{d)} Diabiatic evolution for $\Delta = J/2$. Turning the spins sufficiently slowly still results in a quantum spin helix, $\phi(t) = -0.3 \new{J t /\hbar}$.
}
\end{figure*}

Quantum spin slippage may seem odd on the first sight because $\langle \vec{S} \rangle^2$ cannot vanish for an isolated spin. Here, however, several spins entangle to let the magnitude of $\langle \vec{S} \rangle$ vanish, a situation similar to \new{a} \cancel{the} spin-singlet \cancel{state}.
The generic origin of quantum slippage is quantum mechanical avoided crossing of energy levels of the chain due to which the adiabatic time evolution is incapable of reaching energetically higher states. The situation is depicted in \refFig{figTimeEvolvedSpectrumNotSweet} for a chain of $n=7$ spins with $S=\hbar/2$ at $\Delta = 0$. The energetically lowest two states separate from the rest of the spectrum and cannot be reached by adiabatic time evolution. The width of the blue trace indicates the weight of the dynamic state when decomposed into the instantaneous eigenstates. The situation remains unchanged when rotating the last spin with a finite angular velocity. The diabatic dynamics is depicted in \refFig{figTimeEvolvedSpectrumNotSweetDiabatic}.  Instead of developing a spin helix by tunneling through the avoided level crossing, the system disperses into several eigenstates and the spin texture disorders.

\paragraph{Quantum spin helices at sweet spots.}
As well known from solid states physics, symmetries of the Hamiltonian may occur for special values of the parameters such that avoided level crossings close and quantum spin helices can be wound up.
Such \textit{sweet spots} also exist for the quantum spin chains at hand. We determine numerically those values of the Heisenberg anisotropy $\Delta$, for which the energy gap $E_{1}$ between the ground and the first excited state vanishes in dependence \new{on} $S$ and the chain length. 
Notably, when this particular gap closes, most of the other relevant avoided level crossings close as well, see \cite{SM}. 
We depict the dependence of $E_{1}$ on  $\Delta$ for $S\leq \hbar$ in \refFig{figSweetSpotsCatI}. Cases with $S>\hbar$ are addressed in the Supplemental Material \cite{SM}.
The sweet spots fall into two categories. The first one is universal in $S$ and, to begin with, comprises the values $\Delta = J \cos(\pi/(n-1))$.
Remarkably, \cancel{this angle,} $\pi/(n-1)$, is exactly the averaged twisting angle of adjacent spins. Additionally, for $S= \hbar/2$, the first category includes the values $\Delta = J \cos(\pi/m)$, where $m$ may take any odd integer value smaller than $n$.
\new{
We not that, for infinite spin chains, there are infinitely many sweet spots, which nicely agrees with the gapless ground state of infinite spin-{\small $1/2$} chains in the planar regime \cite{GiamarchiBook2003}.
}
\cancel{In this respect,}  
The sweet spot $\Delta = J/2$ stands out as being independent of the length for chains with $S=\hbar / 2$.
The second category comprises all other sweet spots, which generally depend on both $S$ and $n$ and are tabularized in \cite{SM}. 
It is worth noting that, here, all spin-{\small $1/2$} chains are integrable \cite{RabsonNarozhnyMillis2004CrossoverFromPoissonToWignerDysonLevelStatisticsInSpinChainsWithIntegrabilityBreaking,MurganSilverthorn2015TheSolutionOfAnOpenXXZChainWithArbitrarySpinRevisited}, while all discussed chains with a larger spin quantum number are not \cite{KirillovReshetikhin1987ExactSolutionOfTheIntegrableXXZHeisenbergModelWithArbitrarySpin,Jimbo1985AqDifferenceAnalogueOfUgAndTheYangBaxterEquation,MezincescuNepomechieRittenberg2012BetheAnsatzSolutionOfTheFateevZamolodchikovQuantumSpinChainWithBoundaryTerms}. We corroborate this by a level spacing analysis \cite{BerryTabor1977LevelClusteringInTheRegularSpectrum,PoilblancZimanBellissardMilaMontambaux1993PoissonVSGOEInIntegrableAndNonintegrableQuantumHamiltonians,HsuDAuriac1993LevelRepulsionInIntegrableAndAlmostIntegrableQuantumSpinModels} that indicates a symmetry related origin of the sweet spots, see \cite{SM}.
\begin{figure}
\includegraphics[width = 0.9 \linewidth]{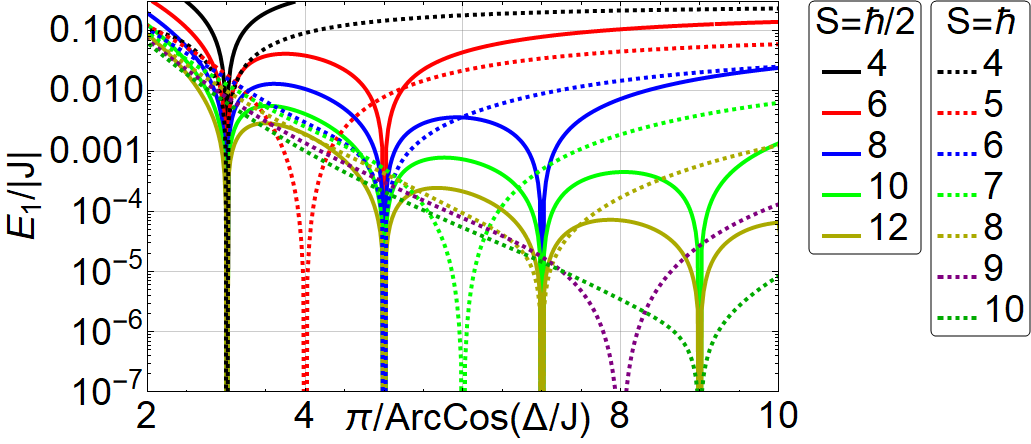} 
\caption{\label{figSweetSpotsCatI}%
Energy gap $E_1$ between the ground state and the first excited state at a twisting angle $\phi = \pi$ in dependence on $\Delta$ for $S \leq \hbar$. The energy gap vanishes (numerically) exactly at the sweet spots $\Delta = J \cos(\pi/m)$, with odd $m<n$ for $S=\hbar/2$ and $m=n-1$ for $S=\hbar$.
}
\end{figure}

The impact of the sweet spots on the dynamic winding-up of a helix is shown in \refFig{figTimeEvolvedSpectrumSweet}. Here, all relevant avoided level crossings for $n=7$, $S=\hbar/2$ close for $\Delta =  J/2$.
An adiabatic or a sufficiently slow diabiatic time evolution is consequently able to wind up a quantum spin helix.
We depict this in \refFig{figTimeEvolvedSpectrumSweet} and \refFig{figTimeEvolvedSpectrumSweetDiabatic}, where the quantum spin helix state climbs up in energy and remains helical without premature slippage.

\begin{figure}
\includegraphics[width = 0.8 \linewidth]{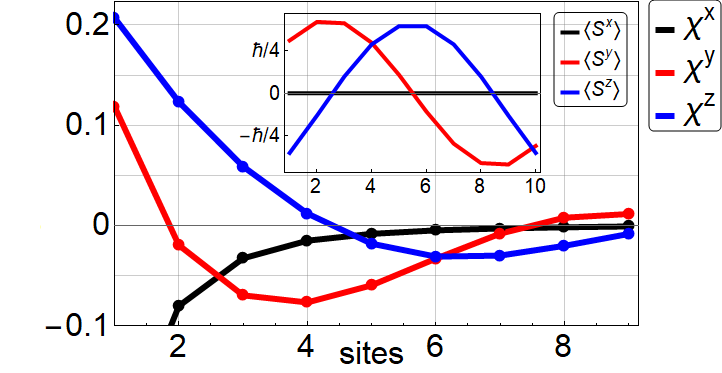}%
\caption{\label{figSpinSpinCorrelations}
Spin-spin correlation in units of $\hbar^2/4$ for $S=\hbar/2$, $\Delta = J/2$, and $n=12$ at a helical state of one full twist (spin expectation values shown in inset). The spins are non-locally entangled.}
\end{figure}

Interestingly, the first category of sweet spots contains the mentioned ones for spin-{\small$1/2$} chains with boundary dissipation \cite{Popkov2016PhysRevA,Popkov2017JPhysA}, which are associated to quantum spin helices formed by a pure product state of local spin-{\small$1/2$} states
$|\Psi\rangle \propto	 \bigotimes_{k=0}^{n-1} \left(e^{-\I \varphi k /2},e^{\I \varphi k /2 }\right)^T$.
The nature of the dynamically constructed quantum helix, here, however, is non-trivial: These quantum helices are non-locally entangled. %
This can be seen from the spin-spin correlation 
$\chi^{\lambda}(s) = \langle S^{\lambda}_{1} S^{\lambda}_{1+s} \rangle - \langle S^{\lambda}_{1}\rangle \langle S^{\lambda}_{1+s}\rangle$ where $\lambda \in \left\{ x,y,z \right\}$, which vanishes for local product states. The result for $S=\hbar/2$, $n=12$ in a helical state of one full twist is shown in \refFig{figSpinSpinCorrelations} and is clearly non-zero for a finite range of spins along the chain.
These helical states are pure eigenstates, which potentially easily decay by external perturbations. \new{Yet}, we find a vast insensitivity against local parametric magnetic fluctuations at the sweet spots in first-order perturbation theory \cite{SM}.

The sweet spots require fine-tuning of the Hamiltonian \cancel{parameters} in order to realize a quantum spin helix. In the following, we present two less restrictive options how the avoided level crossings can be overcome and the slippage angle of a quantum spin chain be increased. The first one is the use of different dynamic Landau-Zener protocols in the vicinity of the sweet spots, while the second one is the coupling to a magnetic environment.

\paragraph{Dynamic Landau-Zener protocols. }
In the vicinity of the sweet spots, $\Delta - \epsilon$ matches the value of a sweet spot for a small $\epsilon$. The relevant level crossings do not vanish in this case, but remain small. The transition is therefore well approximated by a two level Hamiltonian %
\begin{align}
	\mathcal{H}_{\text{LZ}}(\phi) =E_\phi\sigma_0 + \frac{E_{\text{LZ}}}{2} \sigma_x + \frac{v_{\text{LZ}} (\phi+\delta \phi_{\text{LZ}})}{2} \sigma_z\, ,
\end{align}
with the energy gap $E_{\text{LZ}}$, the coupling $v_{\text{LZ}}$, and $\delta \phi_{\text{LZ}}$, which describes a possible shift of the level crossing. Furthermore, $\sigma_i$ are the Pauli matrices and $E_{\phi}$ is the 
energetic background.
For instance, the first such transition of a chain with $S= \hbar/2$ and $n=7$ happens at a twisting angle $\phi = \frac{4}{3} \pi$ and is characterized by $E_{\text{LZ}}/\epsilon = -0.634$, $v_{\text{LZ}} /J = 0.990$, and $\delta \phi_{\text{LZ}} J/\epsilon =0.252$. 
These values, can be applied to known protocols that perfectly overcome the avoided level crossing. 
One is the infinitely fast Landau-Zener transition \cite{GiacomoNikitin2005TheMajoranaFormulaAndTheLandauZenerStueckelbergTreatment}, 
another one is a $\pi$-pulse in a resonant Rabi cycle, 
and a third one is an adiabatic, or piecewise adiabatic frequency chirp \cite{Zhdanovich..Millner2008PopulationTransferBetween2QuantumStatesByPiecewiseChirpingOfFemtosecondPulsesTAndE}.
More involved protocols are also feasible \cite{Bernes2013AnalyticallySolvableTwoLevelQuantumSystemsAndLZInterferometry,JhaRostovtsev2010AnalyticalSolutionsForATwoLevelSystemDrivenByAClassOfChirpedPulses}.
These techniques are not necessarily connected to optical methods in our setup. They merely correspond to being able to rotate the last spin by specific angles.
\cancel{
Concretely, when trying to wind up a quantum spin helix, we suggest to first rotate adiabatically into the vicinity of an avoided level crossing, 
switch to one of the mentioned time-dependent protocols, 
and subsequently continue with an adiabatic revolution.}

\begin{figure}
\includegraphics[width = 0.73 \linewidth]{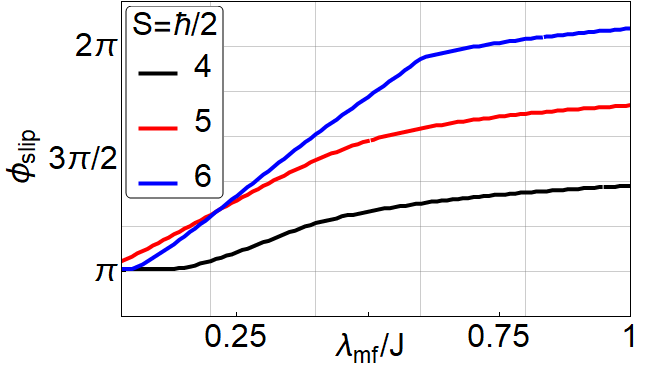}%
\caption{\label{figMeanFieldStabilization} Stabilization of quantum spin helices by a local dissipative bath.
Slippage angle in mean-field approximation for $S=\hbar/2$, $\Delta = 0$ for different chain lengths. 
}
\end{figure}

\paragraph{Dissipative magnetic environment. } 
As depicted in \refFig{figUpWindingOfSpinSpirals}, quantum spin slippage is connected to vanishing local spin expectation values. 
A coupling to a magnetic environment effectively partially measures the quantum spins, gradually turning them into classical spins. By this, the magnitude of the spin expectation vector is stabilized. 
On the other hand, the environment opens additional decay channels  and tries to relax the spins. 
To figure out which effect is dominant, we consider a coupling to local bosonic magnetic fluctuations described by the coupling Hamiltonian  
\begin{align}
\mathcal{H}_{\text{fluct}} &=\alpha \sum_{i=1}^n \sum_{\lambda \in \{x,y,z\}}\sum_\kappa 
{S_{i}^\lambda} \left({b^{\lambda}_{i,\kappa}} + {b^{\lambda \, \dagger}_{i,\kappa}} \right). 
\end{align}
Performing second order Keldysh formalism in $\alpha$ yields a local magnetic backaction of the form $\sum_{i=1}^n \int d\tau \chi(\tau) \vec{S}^I_i(0) \vec{S}^I_i(-\tau)$, where the superscript $I$ indicates the interaction picture. Within mean-field theory, and assuming a stationary state, we arrive at the effective mean-field Hamiltonian (see \cite{SM} for details) 
\begin{align}
\label{eqnHamiltonianMeanField}
\mathcal{H}_{\text{mf}} &=  \lambda_{\text{mf}} \, J  \sum_{i=1}^n \left( 2 \vec{S}_i - \langle \vec{S}_i \rangle\right) \langle \vec{S}_i \rangle,
\end{align}
where $\lambda_{\text{mf}}$ is a real constant.
Thus, the bosonic magnetic fluctuations generate self-stabilizing local Zeeman fields that suppress quantum spin slippage. 
In \refFig{figMeanFieldStabilization}, we depict the slippage angle in dependence on $\lambda_{\text{mf}}$ for  different chain lengths and $S=\hbar/2$. The slippage angle increases with the chain length but eventually saturates in dependence on the dissipation strength. This saturation corresponds to the inability of a classical spin chain to be twisted more than a certain amount before it relaxes \cite{VedmedenkoAltwein2014TopologicallyProtectedMagneticHelixForAllSpinBasedApplications}. We expect the behavior for larger spin quantum numbers, which behave more like classical spins, to be qualitatively the same. Increasing the temperature of the magnetic environment generally increases the mean-field coupling $\lambda_{\text{mf}}$ as well. The spin-spin correlation along the chain, cf. \refFig{figSpinSpinCorrelations}, decreases in $\lambda_{\text{mf}}$, such that at $ |\lambda_{\text{mf}}| \to \infty$ the spiral state \new{recovers} the non-entangled local product states of
\citeRef{Popkov2017PhysRevA}. 

\paragraph{Quantum computing with spin slippage states. } 
Next, we show how the quantum spin chains can be used for spin-chain-based quantum computing.
From \refFig{figTimeEvolvedSpectrumNotSweet}, we observe that the adiabatic evolution away from the sweet spots does not return to the ground state after a revolution by $2\pi$, but only after a revolution by $4 \pi$. The exact level crossing of the first and the second level at an angle of $\pi$, which is needed for this behavior,  is generic and independent of $\Delta$ for the case of half-integer spins and an odd length.
The two crossing levels are Kramers partners, i.e., they are connected by an antiunitary symmetry $\mathcal{A}$ that squares to $-1$.
We find that $\mathcal{A} = A\mathcal{K}$, where 
\begin{align}
A &= T_{\left.j \Leftrightarrow n-j+1 \right.} e^{\I \pi \sum_{k=1}^n S_k^y}
\end{align}
is the unitary transformation representing a rotation of all spins around the $y$-axis by an angle of $\pi$ and the subsequent exchange of the $j^{\text{th}}$ and the $(n-j+1)^{\text{th}}$ spin, denoted by the permutation matrix $T_{\left.j \Leftrightarrow n-j+1 \right.}$. %
Moreover, $\mathcal{K}$ is complex conjugation and we use the standard representation, where $S^y$ is a purely imaginary matrix. 
Squaring $\mathcal{A}$ yields
$\mathcal{A}^2 = (-1)^{2 S n}$, 
which establishes the desired Kramers degeneracy exactly for half-integer spins and chains of odd length. 
The result is independent of $J$ and $\Delta$ %
and hence also applies to antiferromagnetic chains and to the axial regime $|\Delta| \geq |J|$.
The first excited state can, in conjunction with the ground state, be used as an energetically split qubit.
The adiabatic evolution of a full revolution $\phi = 2 \pi$ realizes the unitary quantum gate 
$U_{\pi} = \I \sigma_x. $
After two full revolutions of the last spin, the Berry phase is $\pi$, hence, still nontrivial. 
The protected degeneracy of the ground and first excited state can also be interpreted physically: For $\Delta = J$ (XXX model), one full revolution adiabatically pumps a single magnetic excitation into the chain that carries a total spin of $\hbar/2$, independent of the spin quantum number of the chain. The qubit states may thus be distinguished by measuring the total magnetization or the magnetoresistance. 

\paragraph{Conclusions. } 
The dynamic winding-up of a quantum spin helix shows many non-trivial features as compared to its classical counterpart. The topological protection of the classical helix is generally overcome by tunneling, and quantum spin slippage occurs \new{in chains of finite length}. Only at sweet spots of the axial Heisenberg anisotropy, classical topological protection is restored. Interestingly, topologically protected quantum helices are not formed only by product states of local spin states, but are instead entangled. Quantum spin slippage can be avoided dynamically by Landau-Zener protocols or by a magnetic environment. 
Finally, the first quantum slippage state, in the case of half-integer spins and chains of even length, forms a Kramers symmetry-protected qubit with the ground state,  
which is well-separated from the rest of the spectrum.
The resulting protected adiabatic qubit can be used for  adiabatic quantum computing by having a non-trivial Berry connection.
Arrays of spin chains could potentially be used for universal adiabatic quantum computing.

\paragraph{Acknowledgments. } 
We thank  
Peter Nalbach, Martin Stier, and Elena Vedmedenko for stimulating discussions. We acknowledge support by the Deutsche Forschungsgemeinschaft (DFG, German Research Foundation) - 420120155.

%

\clearpage

\appendix
\renewcommand{\thefigure}{SM-\arabic{figure}}
\renewcommand{\thetable}{SM-\Roman{table}}
\onecolumngrid

\section{Supplemental Material}

\paragraph{
In this Supplemental Material, we analyze the avoided level crossings/exact level crossings at the sweet spots.
We furthermore derive the mean-field approximation of the magnetic bosonic bath, determine the stability of the helical states against external, parametric perturbations and present a level spacing analysis of the spin chains at hand.
}

\subsection{Sweet spot analysis}
\begin{figure*}[b]
\subfloat[\label{figSweetSpotAnalysisPlotNormal}]
{
\includegraphics[height = 0.217 \linewidth]{piGapPlotNormal18.png} 
}
\subfloat[\label{figSweetSpotAnalysisPlotUnNormal}]
{
\includegraphics[height = 0.217 \linewidth]{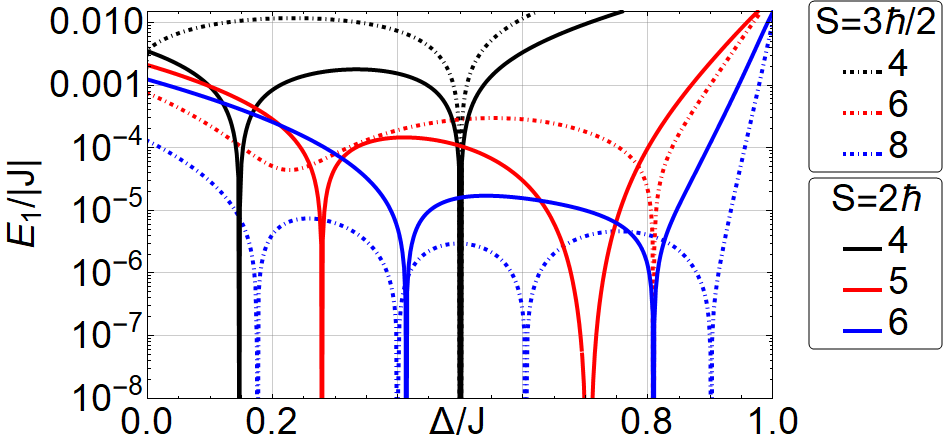} 
}
\caption{\label{figSweetSpotAnalysis}%
Sweet spots of the axial Heisenberg anisotropy $\Delta$, i.e., where the avoided level crossings partially vanish and quantum spin helices are possible.
Shown is the energy gap $E_1$ between the ground state and the first excited state at a twisting angle $\phi = \pi$ in dependence on $\Delta$. \textbf{a)} For $S\leq \hbar$, The energy gap vanishes (numerically) exactly for $\Delta = J \cos(\pi/(n-1))$, where $\pi/(n-1)$ is exactly the twisting of adjacent spins. \textbf{b)} For $S>\hbar$, additional sweet spots appear that are not of the described functional form, see \refTab{tabSweetSpots}.
}
\end{figure*}

At the sweet spots, cf. \refTab{tabSweetSpots},
the ground state and the first excited state become degenerate at a twisting angle of $\phi = \pi$. Quantum spin slippage is suppressed, and the ability to wind up a spin helix is extended over $\phi = \pi$. 
An overview of the energy separation $E_1$ between the ground state and the first excited state at $\phi = \pi$ is given in \refFig{figSweetSpotAnalysis}, where $E_{1}$  is plotted in dependence on the axial Heisenberg anisotropy $\Delta$. The sweet spots stand out by $E_1 \to 0$.
By an adiabatic evolution at one of the sweet spots, the system adiabatically follows the state that initially was the ground state. Of course, at angles $\phi>\pi$ there may be additional level crossings that hinder the system to spiral-up further. 
In general, we find that multiple of these additional avoided level crossings close at the sweet spots as well, however, not all of them --- although even these avoided level crossings are largely suppressed. In \refTab{tabFiniteSizeLevelCrossingsPhi} and \refTab{tabFiniteSizeLevelCrossingsE}, we exemplify this behavior for a chain of different spin quantum numbers. 
The relevant avoided level crossings that do not close at the sweet spots become further suppressed if the length of the chain is increased, see \refTab{tabFiniteSizeLevelCrossingsFiniteSize}. In this sense, the remaining avoided level crossings can be interpreted as finite size effects.
\begin{table}[b]
\begin{tabular}{|c|c|ccccc|}
\hline
cat. &	{\vphantom{$\left(\int^{\int^a}_a\right)$}}$S/\hbar$	&	Sweet spots  $\Delta$ (in units of $J$) \hspace*{-40ex}	&	&	&	&\\
\hline
\hline 
\multirow{2}{*}{$C1$}& {\small$1/2$}&  $\cos(\pi/m)$ with odd $m \leq n-1$	\hspace*{-20ex} 	&  &  &  &  \\
&	$1$ & $\cos(\pi/m)$ with $m=n-1$	\hspace*{-20ex}&  &  &  &  \\
\hline
\hline
\multirow{2}{*}{$C2$} & 	$3/2$ &-1.00000	& -0.0483412 & 0.175242 & 0.399895 & 0.604953 \\
&	$2 $ &	0.146900 & 0.278657 & 0.413477 & 0.497881 & \\ 
\hline
\end{tabular}
\caption{%
\label{tabSweetSpots} Sweet spots $\Delta$ of category $C1$ and $C2$ in dependence on $S$ for $|\Delta|\leq |J|$.
For $S >\hbar$ not all sweet spots are listed, the missing ones are the same as for $S=\hbar$.
}

\end{table}

\begin{table}[t]
\subfloat[\label{tabFiniteSizeLevelCrossingsPhi}]{
\begin{tabular}{|c|ccccc|}
\hline
 $S/\hbar$ & $\phi _1/\pi$ & $\phi _2/\pi$ & $\phi _3/\pi$ & $\phi _4/\pi$ & $\phi _5/\pi$ \\
\hline
\hline
 $1/2$ & $1.000$ & $1.333$ & $1.667$ & $1.797$ & $2.000$ \\
\hline
 $1$ & $1.000$ & $1.078$ & $1.156$ & $1.234$ & $1.339$ \\
\hline
 $3 /2$ & $1.000$ & $1.051$ & $1.102$ & $1.152$ & $1.208$ \\
\hline
 $2 $ & $1.000$ & $1.038$ & $1.076$ & $1.113$ & $1.149$ \\
\hline
\end{tabular}
}
\subfloat[\label{tabFiniteSizeLevelCrossingsE}]{
\begin{tabular}{|c|ccccc|}
\hline
 $S/\hbar$ & $\Delta E_1/J$ & $\Delta E_2/J$ & $\Delta E_3/J$ & $\Delta E_4/J$ & $\Delta E_5/J$ \\
\hline \hline
 $1/2$ & $0$ & $0$ & $0$ & $0$ & $0$ \\
\hline
 $1$ & $0$ & $0$ & $0.007774$ & $0$ & $0.006987$ \\
\hline
 $3/2$ & $0$ & $ 7.154 {\expstyle \times 10^{-6}}$ & $0$ & $0.001288$ & $0.004200$ \\
\hline
 $2$ & $0$ & $0$ & $ 2.886{\expstyle \times 10^{-6}}$ & $0$ & $7.454 {\expstyle \times 10^{-5}}$ \\
\hline
\end{tabular}
}

\subfloat[\label{tabFiniteSizeLevelCrossingsFiniteSize}]{
\begin{tabular}{|c|cccccc|}
\hline
{\vphantom{$\left(\int^{\int^a}_a\right)$}}$S$	& $n=5$ & $6$ & $7$ & $8$ & $9$ & $10$ \\
\hline
\hline
 $\hbar/2$ & $0$ & $0.346$ & $0$ & $0.1955$ & $0$ & $0.1253$ \\
\hline 
$\hbar$ & $0.1213$ & $0.1066$ & $0.0925$ & $0.08061$ & $0.07101$ & $0.06323$ \\
\hline
\end{tabular}
}
\caption{\label{tabFiniteSizeLevelCrossings}%
Analysis of the first Landau-Zener transitions/avoided level crossings at the sweet spots $\Delta = J/2$ for $S=\hbar/2$ and  $\Delta = \cos\left(\pi/(n-1)\right) J$ for $S\geq\hbar$. 
\textbf{a)} Angle $\phi_i$ and \textbf{b)} gap size $\Delta E_i$ of the $i^{\text{th}}$ avoided or exact level crossings for a quantum spin chain of length $n=7$ for different spin quantum numbers $S$ to be overcome when winding up a spin helix. \textbf{c)} Scaling of the energy gap at a twisting angle $\phi=\pi$ for chains of different length $n$. The gaps monotonically decrease in $n$ and decrease on average in $S$. Spin-{\small$1/2$} chains stand out by only showing exact level crossings.
}
\end{table}

\section{Mean-field approximation of the magnetic environment}

In this section, we derive the mean-field Hamiltonian used in the main text that represents the influence of a bosonic magnetic environment on the spin chains.
The full Hamiltonian is
\begin{align}
\mathcal{H} &= \mathcal{H}_{\text{chain}} + \mathcal{H}_{\text{bath}} + \mathcal{H}_{\text{fluct}},
\\
\mathcal{H}_{\text{bath}} &=  \sum_{i=1}^n \sum_{\lambda \in \{x,y,z\}}\sum_\kappa \epsilon_{i,\kappa}^\lambda {b^{\lambda \, \dagger}_{i,\kappa}} {b^{\lambda}_{i,\kappa}},
\\
\mathcal{H}_{\text{fluct}} &=\alpha \sum_{i=1}^n \sum_{\lambda \in \{x,y,z\}}\sum_\kappa 
{S_{i}^\lambda} \left({b^{\lambda}_{i,\kappa}} + {b^{\lambda \, \dagger}_{i,\kappa}} \right),
\end{align}
where $\epsilon_{i,\kappa}^\lambda$ encodes the spectral density of the environmental fluctuations and $\alpha$ is a real coupling constant.
We now derive an effective Hamiltonian for the chain in the same spirit as the Ruderman-Kittel-Kasuya-Yosida (RKKY) Hamiltonian. The derivation follows the lines of \citeRef{Posske2016PhdThesis}.
Additionally, we perform a mean-field decoupling in between.
In the second-order real-time Keldysh formlism \cite{Rammer2007QuantumFieldTheoryOfNonEquilibriumStates}, the expression for the expectation value of a time-ordered operator $A$, given in the interaction picture, is 
\begin{align}
\left\langle A \right\rangle = \left\langle A \right\rangle_0 -  \int_c d\tau d\Delta_\tau \sum_{i,j=1}^{n} \sum_{\lambda,\lambda^\prime} \alpha^2 \chi^{\lambda, \lambda^\prime}(\Delta_\tau) \left\langle \mathcal{T} S_i^{\lambda}(\tau) S^{\lambda^\prime}_j(\tau-\Delta_\tau) A \right\rangle_0 +  \mathcal{O}\left(\alpha^4\right),
\end{align} 
with the susceptibility
\begin{align}
\chi^{\lambda, \lambda^\prime}_{i,j}(\Delta_\tau) &= \I \sum_{\kappa,\kappa^\prime}\left\langle  \mathcal{T} \left({b^{\lambda}_{i,\kappa}} + {b^{\lambda \, \dagger}_{i,\kappa}} \right) \left({b^{\lambda^\prime}_{j,\kappa^\prime}}\left(\Delta_\tau\right) + {b^{\lambda^\prime \, \dagger}_{j,\kappa^\prime}}\left(\Delta_\tau\right) \right)  \right\rangle_0 
\nonumber \\
&= \I \delta_{\lambda,\lambda^\prime} \delta_{i,j} \sum_\kappa  \left\langle \mathcal{T}\left\{ b^\lambda_{i,\kappa} b^{\lambda \, \dagger}_{i,\kappa}(\Delta_\tau) +  b^{\lambda \, \dagger}_{i,\kappa} b^{\lambda}_{i,\kappa}(\Delta_\tau) \right\} \right\rangle
\nonumber \\
&=: \I \delta_{\lambda,\lambda^\prime} \delta_{i,j} \chi(\Delta_\tau).
\end{align}
Here, $\langle \dots \rangle$ denotes the expectation value with respect to the full Hamiltonian, $\langle \dots \rangle_0$, the expectation value with respect to the bare Hamiltonian $\mathcal{H}|_{\alpha = 0}$, $\int_c d\tau \dots$ denotes integration over the Keldysh contour, $\mathcal{T}$ is the Keldysh contour time ordering, and all operators are represented in the interaction picture, where $\mathcal{H}_{\text{fluct}}$ is the interaction Hamiltonian.
Hence, we obtain
\begin{align}
\left\langle A \right\rangle = \left\langle A \right\rangle_0 -  \int_c d\tau d\Delta_\tau \sum_{i} \sum_{\lambda} \alpha^2 \chi(\delta_\tau) \left\langle \mathcal{T} S_i^{\lambda}(\tau) S^{\lambda}_i(\tau-\Delta_\tau) A \right\rangle_0 +  \mathcal{O}\left(\alpha^4\right).
\end{align}
In order to estimate the second order term, we perform a mean-field decoupling of the spin operators. To this end, we rewrite, as usual, $S^\lambda_i 
 = \delta S^\lambda_i + \langle S_i^\lambda\rangle $ with 
$ \delta S^\lambda_i = S^\lambda_i - \langle S_i^\lambda \rangle$ and neglect terms of the form $\left(\delta S_i^\lambda\right)^2$. We then obtain
\begin{align}
\label{eqnMeanFieldApproximatedSpins}
S_i^{\lambda}(\tau) S^{\lambda}_i(\tau-\Delta_\tau)  &\stackrel{\text{mf}}{=} \delta S_i^\lambda(\tau) \langle S^\lambda_i(\tau-\Delta_\tau) \rangle + \delta S_i^\lambda(\tau- \Delta_\tau) \langle S^\lambda_i(\tau)\rangle + \langle S^\lambda_i(\tau-\Delta_\tau) \rangle \langle S^\lambda_i(\tau)\rangle
\nonumber \\
&\approx 
 \left(2 S_i^\lambda(\tau) - \langle S_i^\lambda(\tau)  \rangle \right) \langle S_i^\lambda(\tau) \rangle,
\end{align}
where $ \stackrel{\text{mf}}{=}$ stands for the above mentioned mean-field approximation.
We furthermore assume in the second line of \refEq{eqnMeanFieldApproximatedSpins} that the dynamics of the spin system is considerably slower than the one of the bosonic fluctuations, such that we can forget about the contour time $\Delta_\tau$ in the final result. This corresponds to the Markov approximation in dissipative systems.
In total, we arrive at 
\begin{align}
\label{eqnFinalMeanFieldExpression}
\left\langle A \right\rangle \stackrel{\text{mf}}{=} \left\langle A \right\rangle_0 -  \int_c d\tau  \sum_{i,\lambda} \alpha^2 \chi \left\langle \mathcal{T}  \underline{\left(2 S_i^\lambda(\tau) - \langle S_i^\lambda(\tau)  \rangle \right) \langle S_i^\lambda(\tau) \rangle} A \right\rangle_0 +  \mathcal{O}\left(\alpha^4\right),
\end{align} 
with the final form of the susceptibility
\begin{align}
\chi = \sum_\kappa \int_c d\Delta_\tau \left\langle \mathcal{T}\left\{ b^\lambda_{i,\kappa} b^{\lambda \, \dagger}_{i,\kappa}(\Delta_\tau) +  b^{\lambda \, \dagger}_{i,\kappa} b^{\lambda}_{i,\kappa}(\Delta_\tau) \right\} \right\rangle,
\end{align}
which is a real number. 
The result shows that there is exactly one Hamiltonian (underscored in \refEq{eqnFinalMeanFieldExpression}) that only acts on the spin space and results in (up to second order in $\alpha$) the same expectation values for all operators $A$. This is the mean-field Hamiltonian of the main text:
\begin{align}
\mathcal{H}_{\text{mf}}^{\text{total}} &=  H_{\text{chain}} + \mathcal{H}_{\text{mf}}
\\
\mathcal{H}_{\text{mf}} &= \lambda_{\text{mf}} \, J \sum_{i=1}^n \sum_{\lambda \in \{x,y,z\}}  \left(2 S_i^\lambda - \langle S_i^\lambda  \rangle \right) \langle S_i ^\lambda \rangle,
\end{align}
with 
\begin{align}
\lambda_{\text{mf}} &= \alpha^2 \chi / J.
\end{align}

\section{Stability analysis of the helical states}

\begin{figure*}
\centering
 \includegraphics[width = 0.6 \linewidth]{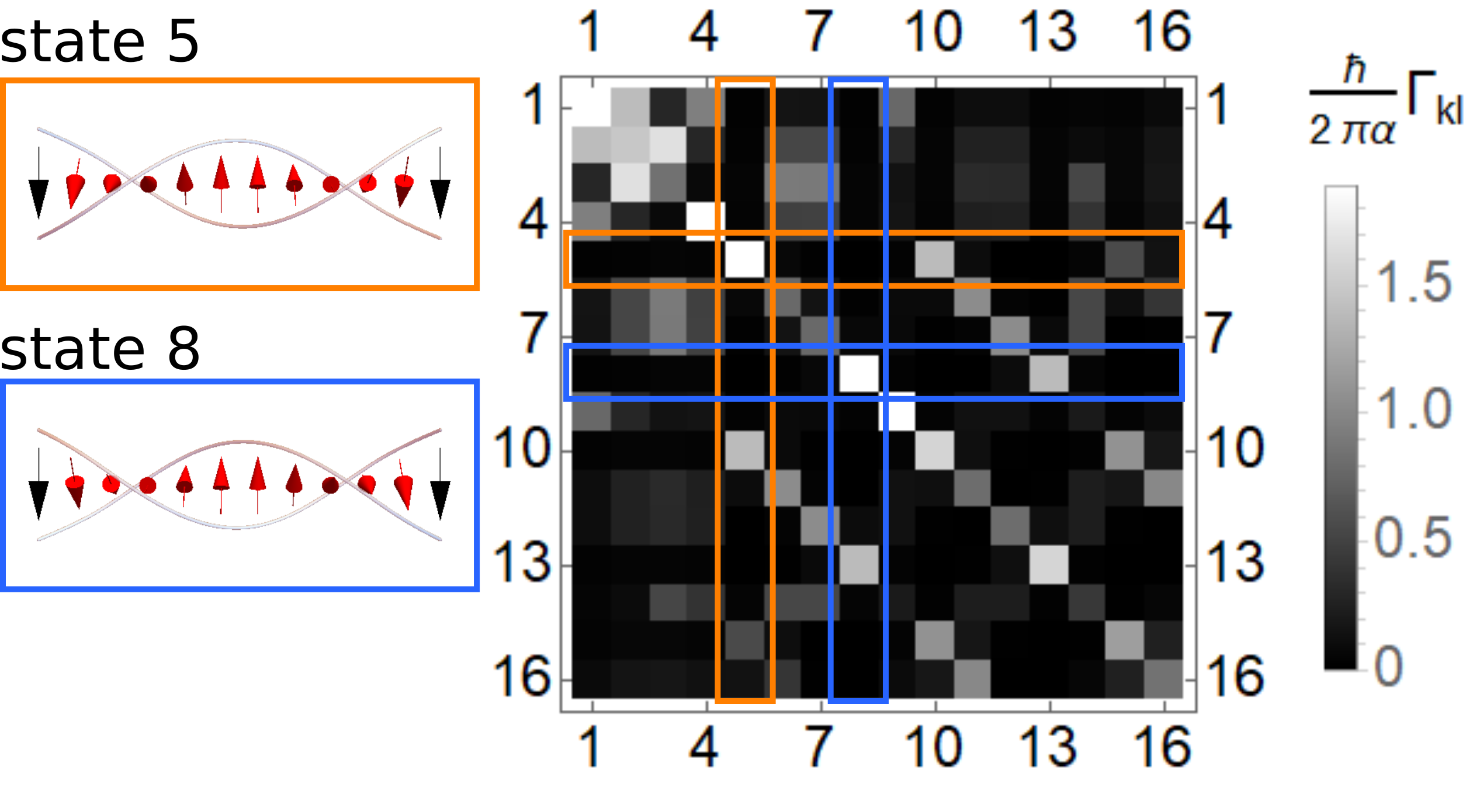}
 \caption{\label{figTransitionRatesHelices} Upper boundaries $\Gamma_{kl}$ for the transition rate between the eigenstates $|\Psi_k\rangle$ and $|\Psi_l\rangle$ (see text) for a chain of $10$ quantum spin-{\small$1/2$} at the sweet spot $\Delta = J/2$ that is perturbed by local magnetic fluctuations.
 The energetically lowest helical states (number $5$ and $8$) show almost vanishing transition rates to states with a trivial winding number (not helical). All states are ordered by energy, state 1 is the ground state. The states $5$,$6$,$7$,$8$ are degenerate at a twisting angle of $\phi = 0$. For convenience, this degeneracy is lifted by using a small twisting angle of $\phi = 10^{-3}$.}
\end{figure*}

The helical states at the sweet spots are pure eigenstates of the Hamiltonian with a finite energy difference to the ground state. Hence, once prepared, one would expect that external perturbations induce transitions of the helical state to the ground state, which reduces the applicability of the helical states for practical purposes.
Interestingly, it turns out that the helical states are more stable than expected. In order to show this, we perform first-order perturbation theory in fluctuating local external magnetic fields by Fermi's golden rule.
As external perturbations, we consider all local magnetic fluctuations coupling to $S^x_i$, $S^y_i$, and $S^z_i$ for all sites $i$.
The transition rate from one eigenstate $|\Psi_j\rangle$ to another $|\Psi_k\rangle$ is then given by Fermi's golden rule as
\begin{align}
\gamma^{S^\lambda_i}_{j,k} = \left(2\pi \alpha /\hbar\right) |\langle \Psi_j | S^\lambda_{i,k} |\Psi_k \rangle|^2,
\end{align}
if the frequency of the perturbation perfectly matches the energy difference between the two eigenstates. 
Here, $\alpha$ is a dimensionless, small coupling constant, which we assume to be the same for all external perturbations for simplicity.
In order to obtain a good upper bound, $\Gamma_{j,k}$, for the total transition rate of one state to another one, we sum up the individual transition rates.
\begin{align}
\Gamma_{j,k} = \sum_{\lambda,i} \gamma^{S^\lambda_i}_{j,k}.
\end{align}
Intriguingly, we find that the helical magnetic states are almost insensitive to the considered external perturbations at the sweet spots. The matrix $\Gamma$ is shown in \refFig{figTransitionRatesHelices} for a chain of $10$ quantum spin-{\small$1/2$}. While non-helical states, with winding number zero, decay to the ground state (top left), the helical magnetic states have exceedingly small transition rates to states whose winding number is zero. This is shown, e.g., in columns $5$ and $8$ of the matrix, where the helical states show negligibly small interactions with energetically lower states. The only states it couples to by the external perturbations are helix-like states with larger energy. 
The colorscale does not resolve the mentioned small transition rates of a helical state to a trivial one. We find that these rates further vanish for larger spin chains. Hence, for sufficiently long chains, the helical magnetic state is arbitrarily stable in first-order perturbation theory.
The situation changes away from the sweet spots. There, the transition rates of helical states to non-helical ones are not negligible and no finite size effect. Still, helically magnetized states remain to have a reduced transition rate to trivial states compared to the mean transition rate from a non-helical state to other non-helical states. 

\section{Level spacing statistics and integrability analysis of the sweet spots}
The quest for an analytical explanation of the sweet spots naturally arises. The first idea that may come to mind is that the sweet spots are special points in parameter space where the model becomes integrable.
From what we find, however, integrability is not the key factor.  
First of all, all spin-{\small $1/2$} XXZ Heisenberg chains with nearest neighbor interactions are integrable \cite{RabsonNarozhnyMillis2004CrossoverFromPoissonToWignerDysonLevelStatisticsInSpinChainsWithIntegrabilityBreaking}, being solvable by the Bethe ansatz. This property remains unchanged by the employed boundary terms \cite{MurganSilverthorn2015TheSolutionOfAnOpenXXZChainWithArbitrarySpinRevisited}.
Likewise, for chains with a larger spin quantum number, all models we consider are \textit{not} integrable.
Although, for special parameter values, there exist integrable spin chains with a spin quantum number larger than {\small $1/2$} \cite{KirillovReshetikhin1987ExactSolutionOfTheIntegrableXXZHeisenbergModelWithArbitrarySpin,Jimbo1985AqDifferenceAnalogueOfUgAndTheYangBaxterEquation}, the models we consider are not integrable for $S\geq\hbar$ because of missing axial anisotropies, i.e., terms of the form $({S_i^\lambda})^2$ \cite{MezincescuNepomechieRittenberg2012BetheAnsatzSolutionOfTheFateevZamolodchikovQuantumSpinChainWithBoundaryTerms}. 
Hence, from the point of view of integrability, the sweet spots do not stand out particularly from the rest of the parameters space.

Additionally, because the Hamiltonian is time-dependent, the question arises at which time the integrability of the system is most important. In this regard, it is worth noting that boundary conditions can change the integrability of a system but fail to do so for our boundary conditions, i.e., all our spin-{\small $1/2$} chains are integrable, while all our $S\geq\hbar$ chains are not integrable at all times. For simplicity, we therefore restrict the following analysis to $t=0$, i.e., the magnetic fields at the boundaries point into the same direction.

The level spacing distribution of a Hamiltonian is the probability density function of the energy difference between consecutive eigenvalues of the Hamiltonian.
An important theoretical result is that integrable spin chains follow a Poissonian level spacing distribution $P(\Delta E)\propto e^{-\Delta E \alpha}$, where $\Delta E$ is the level spacing and $\alpha$ is a real number \cite{BerryTabor1977LevelClusteringInTheRegularSpectrum}. Non-integrable spin chains, however, cross over to the eigenvalue statistics of the Gaussian orthogonal ensemble (GOE), i.e., orthogonal matrices with random matrix elements with constant variance. The resulting level spacing distribution is well described by Wigner's surmise $P(\Delta E) \propto \Delta E e^{-\Delta E \alpha}$ \cite{HsuDAuriac1993LevelRepulsionInIntegrableAndAlmostIntegrableQuantumSpinModels,PoilblancZimanBellissardMilaMontambaux1993PoissonVSGOEInIntegrableAndNonintegrableQuantumHamiltonians}.
There is one important prerequisite for these statements to hold rigorously, which is that the statistics from different quantum numbers (if the system has good quantum numbers) need to be counted separately. In the case of open XXZ Heisenberg chains, these are the total spin $\vec{S}_\text{Total} = \sum_i \vec{S}_i$, the total spin-$z$ component $S^z_\text{Total}$, and (in case of symmetric boundary conditions) the parity \cite{HsuDAuriac1993LevelRepulsionInIntegrableAndAlmostIntegrableQuantumSpinModels}.
Furthermore, the level spacing statistics needs to be renormalized by the level density \cite{BerryTabor1977LevelClusteringInTheRegularSpectrum,PoilblancZimanBellissardMilaMontambaux1993PoissonVSGOEInIntegrableAndNonintegrableQuantumHamiltonians,HsuDAuriac1993LevelRepulsionInIntegrableAndAlmostIntegrableQuantumSpinModels}. 

\begin{figure*}
\centering
\parbox{0.5\linewidth}{\raggedright \textbf{a)}\\
\includegraphics[width = 1 \linewidth]{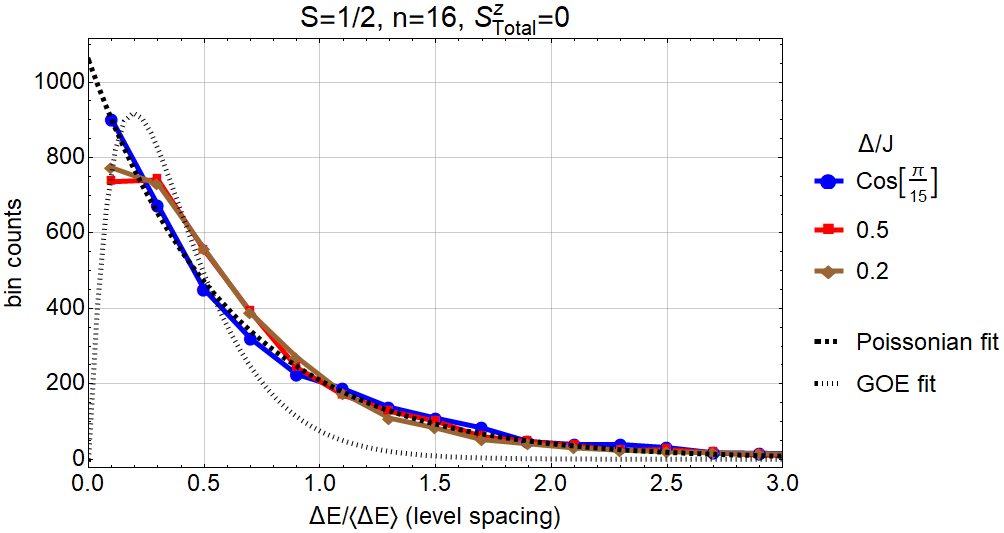}}%
\parbox{0.5\linewidth}{\raggedright \textbf{b)}\\
\includegraphics[width = 1 \linewidth]{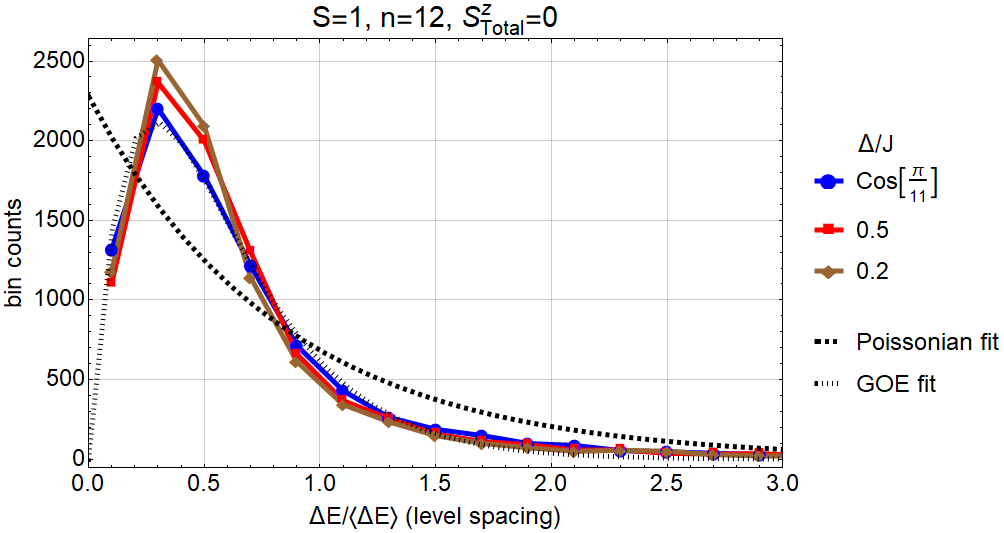}}
\caption{\label{figLevelSpacingDistributions} Level spacing statistics (energy difference distribution of adjacent eigenvalues) of spin chains in dependence on the Heisenberg anisotropy $\Delta$ restricted to $S^z_\text{Total} = 0$. \textbf{a)} A spin-{\small $1/2$} chain with $16$ sites. The distribution is almost Poissonian (dashed curve), signifying an integrable system. The sweet spots (blue and red graph) do not stand out significantly.
\textbf{b)} A spin-{\small$1$} chain with $12$ sites. The distribution follows the GOE statistics (see text), signifying a non-integrable system. The sweet spot value (blue graph) does not stand out significantly from the other values.} 
\end{figure*}

For our considerations, it suffices to restrict ourselves to the level spacing statistics of the spin sector $S^z_\text{Total} = 0$.
The level spacing distributions depicted in \refFig{figLevelSpacingDistributions} confirm that our spin-{\small $1/2$} chains are all integrable while our \mbox{spin-{\small $1$}} chains are not integrable. There is no significant deviation of the level spacing distribution from the Poissonian statistics for spin-{\small$1/2$} chains and no significant deviation from the GOE statistics for spin-{\small $1$} chains.
The depicted level statistics especially show that the discovered sweet spots do not stand out from the rest of the parameter values if only the integrability of the system is regarded.

\begin{figure*}
\centering
\includegraphics[width = 0.5 \linewidth]{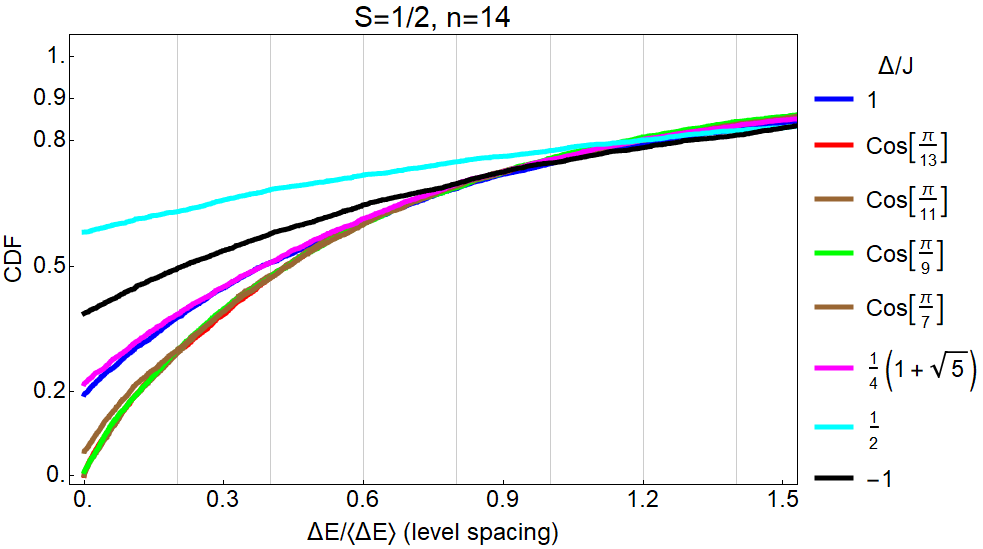}
\caption{\label{figSymmetryProtectionLevelSpacingDistributions}CDF (cumulated  distribution function) of the level spacings of a spin-{\small $1/2$} chain with $14$ sites. In contrast to \refFig{figLevelSpacingDistributions}, $S^z_{\text{Total}}$ is not restricted. The offset at vanishing energy spacing quantifies the amount of degeneracy. At the sweet spot, the amount of degeneracy is increased, with most degeneracy at the dominant sweet spot $\Delta = J/2$. For this value, the chain is more degenerate than for all other values including the $SU(2)$ symmetric case $\Delta = J$.}  
\end{figure*}

The level spacing distribution nevertheless reveals important information about the origin of the sweet spots. If we do not restrict our analysis to the sector $S^z_{\text{Total}} = 0$ but instead include all eigenvalues into the analysis, the level spacing distribution reveals a strong degree of degeneracy exactly at the sweet spots, indicating the vanishing avoided level crossings. This behavior points towards a correlated energetic alignment of the different spin sectors in order to enable the formation of a helical state.
The features that are induced by symmetries at the sweet spots are reflected by a divergence of the probability distribution function of the level spacings at zero level spacing. In order to depict the symmetry protection, we therefore rely on the cumulative distribution function (CDF) of the energy level spacings.
The CDF for a spin-{\small $1/2$} chain with $14$ sites is shown in \refFig{figSymmetryProtectionLevelSpacingDistributions}, where an increasing amount of degeneracy for decreasing sweet spot values of $\Delta$ becomes apparent. The amount of degeneracy at the dominant sweet spot of $\Delta = J/2$ is even larger than for the $SU(2)$-symmetric XXX model ($\Delta = J$) and the XX(-X) model ($\Delta = -J$).

\end{document}